\documentclass[fleqn,usenatbib]{mnras}

\usepackage{newtxtext}
\usepackage{newtxmath}

\usepackage[T1]{fontenc}
\usepackage{anyfontsize} 

\makeatletter

\patchcmd{\NAT@citex}
  {\@citea\NAT@hyper@{%
     \NAT@nmfmt{\NAT@nm}%
     \hyper@natlinkbreak{\NAT@aysep\NAT@spacechar}{\@citeb\@extra@b@citeb}%
     \NAT@date}}
  {\@citea\NAT@nmfmt{\NAT@nm}%
   \NAT@aysep\NAT@spacechar\NAT@hyper@{\NAT@date}}{}{}

\patchcmd{\NAT@citex}
  {\@citea\NAT@hyper@{%
     \NAT@nmfmt{\NAT@nm}%
     \hyper@natlinkbreak{\NAT@spacechar\NAT@@open\if*#1*\else#1\NAT@spacechar\fi}%
       {\@citeb\@extra@b@citeb}%
     \NAT@date}}
  {\@citea\NAT@nmfmt{\NAT@nm}%
   \NAT@spacechar\NAT@@open\if*#1*\else#1\NAT@spacechar\fi\NAT@hyper@{\NAT@date}}
  {}{}
  
\makeatother

\DeclareRobustCommand{\VAN}[3]{#2}
\let\VANthebibliography\thebibliography
\def\thebibliography{\DeclareRobustCommand{\VAN}[3]{##3}\VANthebibliography}

\usepackage{xcolor}	
\usepackage{graphicx}	
\usepackage{amsmath}	

\usepackage{bm} 
\usepackage{textgreek} 

\usepackage[capitalize]{cleveref} 
\creflabelformat{equation}{#2#1#3} 
\crefname{enumi}{item}{items} 

\usepackage{siunitx} 
\DeclareSIUnit[number-unit-product = ]\percent{\char`\%} 

\usepackage[normalem]{ulem} 
\usepackage{csquotes} 
\definecolor{blackberry}{HTML}{8D1D75}

\newcommand*{\code}[1]{\texttt{#1}} 

\newcommand*{\ngc}[1]{NGC\,#1}
\newcommand*{\m}[1]{M\,#1}

\DeclareSIUnit\parsec{pc}
\DeclareSIUnit\dex{dex}
\DeclareSIUnit\h{\mathnormal{h}}
\DeclareSIUnit\year{yr}
\DeclareSIUnit\years{yrs}
\DeclareSIUnit\arcsec{arcsec}
\DeclareSIUnit\arcmin{arcmin}
\DeclareSIUnit\Msun{M_\odot}
\DeclareSIUnit\Rsun{R_\odot}
\DeclareSIUnit\Lsun{L_\odot}
\DeclareSIUnit\Rvir{\mathnormal{R}_\mathrm{vir}}
\DeclareSIUnit\Rhalf{\mathnormal{R}_{1/2}}
\DeclareSIUnit\erg{erg}
\DeclareSIUnit\angstrom{\text{Å}}

\newcommand*{\Msun}{\ensuremath{\mathrm{M}_\odot}} 
\newcommand*{\Rsun}{\ensuremath{\mathrm{R}_\odot}} 
\newcommand*{\Lsun}{\ensuremath{\mathrm{L}_\odot}} 
\newcommand*{\Rvir}{\ensuremath{R_\mathrm{vir}}} 
\newcommand*{\Rhalf}{\ensuremath{R_{1/2}}} 
\newcommand*{\FeH}{\ensuremath{\mathrm{[Fe/H]}}} 

\title[GC formation times from models]{Globular cluster ages and their relation to high-redshift stellar cluster formation times from different globular cluster models}

\author[L. M. Valenzuela et al.]{
Lucas M. Valenzuela,$^{1}$\thanks{E-mail: lval@usm.lmu.de}
Duncan A. Forbes,$^{2}$
and
Rhea-Silvia Remus$^{1}$
\\
$^{1}$Universitäts-Sternwarte, Fakultät für Physik, Ludwig-Maximilians-Universität München, Scheinerstr. 1, 81679 München, Germany\\
$^{2}$Centre for Astrophysics \& Supercomputing, Swinburne University, Hawthorn, VIC 3122, Australia
}

\date{Accepted XXX. Received YYY; in original form ZZZ}

\pubyear{\the\year{}}

\begin{document}
\label{firstpage}
\pagerange{\pageref{firstpage}--\pageref{lastpage}}
\maketitle

\begin{abstract}
The formation details of globular clusters (GCs) are still poorly understood due to their old ages and the lack of detailed observations of their formation.
A large variety of models for the formation and evolution of GCs have been created to improve our understanding of their origins, based on GC properties observed at $z=0$.
We present the first side-by-side comparison of six current GC formation models with respect to their predictions for the GC ages and formation redshifts in Milky Way (MW)-like galaxies.
We find that all the models are capable of forming most of the surviving GCs at more than \SI{10}{\giga\year} ago, in general agreement with the observation that most GCs are old. However, the measured MW GC ages are still systematically older than those predicted in the galaxies of four of the models.
Investigating the variation of modelled GC age distributions for general MW-mass galaxies, we find that some of the models predict that a significant fraction of MW-mass galaxies would entirely lack a GC population older than \SI{10}{\giga\year}, whereas others predict that all MW-mass galaxies have a significant fraction of old GCs. This will have to be further tested in upcoming surveys, as systems without old GCs in that mass range are currently not known.
Finally, we show that the models predict different formation redshifts for the oldest surviving GCs, highlighting that models currently disagree about whether the recently observed young star clusters at high redshifts could be the progenitors of today's GCs.
\end{abstract}




\section{Introduction}
\label{sec:introduction}

As some of the oldest objects in the Universe, globular clusters (GCs) are ideal objects to study the formation histories of galaxies.
Overall, the details of GC formation and evolution are still poorly understood. While measurements of their ages have large uncertainties, observations of GCs in and beyond the Milky Way (MW) suggest that they are generally old, with ages mostly older than \SI{10}{\giga\year} \citep[e.g.,][]{salaris+97, cohen+98, strader+05, vandenberg+13, usher+19}. Some galaxies have been found to 
contain younger GC populations as well, with ages below \SI{8}{\giga\year} \citep[e.g.,][]{sharina+06} or even below \SI{3}{\giga\year} \citep[e.g.][]{goudfrooij+01:GCprops, sharina+06, chomiuk+08, martocchia+18, sesto+18}, though such systems still contain a significant old GC population \citep[e.g.,][for \ngc{1316}, consistent with models of \SI{14}{\giga\year} old GCs]{goudfrooij+01:merger}, and only lower-mass galaxies like the Magellanic Clouds may host mostly younger GCs, where the Small Magellanic Cloud only has GCs below 8--\SI{9}{\giga\year}, with most lying around 2--\SI{5}{\giga\year} \citep[e.g.,][]{parisi+14}.
Younger GCs have been shown to correlate with wet merger events both from observations and simulations \citep{valenzuela+24:GCs}.

For our own MW, the wealth of data obtained through Gaia measurements in the last decade have uncovered more details of the early MW's history through star and GC phase space properties \citep[e.g.,][]{gaia_collaboration18:dr2_gckinematics, deason&belokurov24}. New structures in this phase space have been identified that likely fell into the MW at early times, such as Gaia-Sausage/Enceladus (GSE; \citealp{belokurov+18, helmi+18}). This has also provided researchers with the means to distinguish in-situ and ex-situ formed GCs \citep[e.g.,][]{belokurov&kravtsov24, chen&gnedin24:MWassembly} or even attempt to associate GCs with their progenitor host galaxy \citep[e.g.,][]{massari+19, myeong+18, forbes20, callingham+22}.

Outside the MW, a colour bimodality of GCs has been observed around more massive galaxies and has been found to correlate with their metallicity, ages, and spatial distribution: blue GCs tend to be metal-poor, older, and are found further out, whereas red GCs are metal-rich, younger, and overall trace the inner light distribution \citep[e.g.,][]{peng+04:GCsII, brodie&strader06, faifer+11, usher+12, pota+13, dolfi+21, kluge+23}. This bimodality has been put into relation with the formation pathways of the GCs, where the blue GCs would tend to have formed ex-situ and were then later accreted onto the main galaxy, whereas red GCs would mostly have formed in-situ \citep[e.g.,][]{li&gnedin14, harris+15, forbes&remus18}. Support for this idea has also been found through observations of the blue GCs following the dark matter (DM) halo measured through X-ray gas emission \citep[e.g.,][]{forte+05, forbes+12:ETGs, kluge+23}.

A matter of current research is how early the first star clusters were able to form. Observations with JWST are now revealing compact young star clusters in high-redshift galaxies at redshifts up to $z=10$ in strongly lensed galaxies \citep[e.g.,][]{vanzella+22:abell2744, vanzella+23, adamo+24, mowla+24}. These measurements indicate that bursty star formation already occurs very early on in the Universe. However, the open question remains how these star clusters evolve with time and whether they could be the progenitors of the GCs we observe today at $z=0$.
Studies of intermediate to high redshift systems like the Sparkler galaxy suggest that GCs formed very early can at least survive to redshifts of $z=1.4$ \citep[e.g.,][]{mowla+22, forbes&romanowsky23}.

Many different theoretical approaches have been formulated for GC formation, with early ideas proposing infalling cold gas fragmentation \citep{fall&rees85} or formation through a gas-rich merger \citep{ashman&zepf92}. Some of the recent proposals are also based on colliding scenarios like dwarf galaxy mergers \citep{lahen+19} or collisions of DM subhalos \citep{madau+20}. Other scenarios include relative streaming velocities between baryons and DM \citep[e.g.,][though it has been put into question whether this channel could reach sufficiently high masses, see \citealp{schauer+21}]{naoz&narayan14, lake+21}, formation within accreting gas filaments \citep{van_donkelaar+23}, accreted nuclear star clusters within dwarfs \citep{gray+24}, or disc fragmentation \citep{mayer+24}.
Additionally, some studies made predictions for JWST observations of massive GCs to constrain the initial mass function \citep{jerabkova+17}. However, as of yet it is still under debate which of these channels are actually responsible for the formation of real GCs and how frequent they are.

To investigate the formation and evolution of GCs, numerous GC models have been developed to improve our understanding concerning GCs. These models are developed to explain GC properties seen at $z=0$, and provide a means to constrain the details and environment of their formation.
In order to achieve a sufficiently large sample of GCs for comparison with present-day GCs and simultaneously avoiding impossibly high computational costs, a variety of methods have been used. Empirical and semi-analytic approaches can be applied to artificially created halo merger trees \citep[e.g.,][]{beasley+02, boylan_kolchin17, el_badry+19, de_lucia+24:GCs} or merger trees extracted from cosmological simulations \citep[e.g.,][]{choksi+18, valenzuela+21, chen&gnedin22, chen&gnedin24:MW_M31}. A generally more computationally costly approach is to include GC recipes as a subgrid model within a hydrodynamical cosmological simulation, which includes the E-MOSAICS project \citep{pfeffer+18:emosaics} and EMP-Pathfinder \citep{reina_campos+22:emppathfinder}.
The models are not only diverse with respect to the implementation of GC formation and evolution, but also with respect to the galaxy formation model: some use the intrinsic properties of the hydrodynamic simulation \citep[e.g.,][]{pfeffer+18:emosaics, reina_campos+22:emppathfinder, doppel+23}, whereas most of the models use an underlying empirical or semi-analytic model for galaxy formation.

Even though such models are created to match some collection of properties at $z=0$, they are highly diverse in their overall approach and implementation.
Despite this diversity, there is a general lack of direct comparison between models in the literature.
As none of the models is currently tuned to match high-redshift observations, they each make predictions with respect to the formation times and formation environments of GCs.
In this work we present the first side-by-side comparison of numerous different GC models, for which we study their predictions for GC ages and the formation redshifts of surviving GCs.

As in the MW the ages of GCs can be measured through the distribution of their individual stars in the colour-magnitude diagram (CMD), they have significantly smaller uncertainties than measurements of GC ages in other galaxies. Additionally, most models have focused on MW-like galaxies or at least included such simulated host galaxies in their analyses.
For these reasons, we perform the GC model comparison for MW-like galaxies and compare the results to the measurements in our own MW.
We aim to present the range of direct predictions regarding the ages of GCs, thereby laying the groundwork for improving on these models in the future.

The paper is structured as follows: In \cref{sec:data}, we introduce the observational data for our MW as well as the GC models that we use in our comparison.
Next, we investigate in \cref{sec:age_distributions} whether the models are capable of reproducing the observational data for the GC age distribution in the MW, followed in \cref{sec:age_distribution_variance} by the predicted GC age distributions for MW-mass galaxies for the models that have been applied to a larger number of such galaxies.
In \cref{sec:redshift_distributions}, we present the predicted formation redshift distributions of surviving GCs and compare these to recent high-redshift observations of young star clusters.
Finally, we summarise and conclude in \cref{sec:conclusion}.

\section{Data}
\label{sec:data}

For this work, we collected GC age data from multiple observational and theoretical studies. The observations are of the GCs in the MW and we selected GC formation models that have been run on MW-like galaxies. In the following, we introduce the data sets and detail which criteria were used to define \emph{MW-like} galaxies for the theoretical studies. If applicable, we also state which selection criteria we apply to the GCs, such as mass or metallicity cuts.
Note that the definition of a galaxy being \enquote{MW-like} or a MW analogue depends on the scientific question and has no single answer \citep[see also][]{pillepich+24}. As a result, the different theoretical studies have also applied several different approaches to selecting MW analogues, some more and some less restricting. In this work, we decided to use the term \emph{MW-like} for any galaxy that was selected by more criteria than just halo or stellar mass, such as it being disc-dominated or having an early formation history, for example. In contrast, \emph{MW-mass} refers to any galaxy having been purely selected based on mass criteria.
For all the plots in this and the subsequent sections we show the ages and equivalent redshift scales at the bottom and top of the panels according to the cosmological parameters $h = 0.6781$ and $\Omega_M = 0.308$.

\subsection{Observations}

\begin{figure}
    \centering
    \includegraphics[width=0.9\columnwidth]{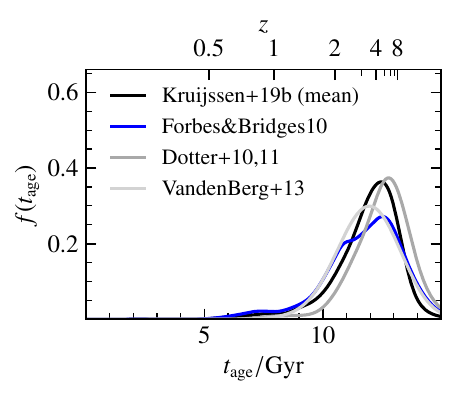}
    \caption{Age distributions of observed GCs in the MW. The ages are taken from \citet{forbes&bridges10}, \citet{dotter+10:acsgcIX, dotter+11}, \citet{vandenberg+13}, and \citet{kruijssen+19:kraken}. The ages from the latter work are the mean ages from the other three studies, which we show as the black distribution. The smooth curves are determined from the sum of normal distributions at the individual age measurements with standard deviations equal to the individual age uncertainties.}
    \label{fig:gc_observations}
\end{figure}

We include three compilations of MW GCs that use resolved CMD data to determine the GC ages, all of which have sample sizes of more than 50~GCs. We took the ages and their uncertainties from \citet{kruijssen+19:kraken} for the following samples, as well as their determined mean GC ages based on the three studies:
\begin{itemize}
    \item \citet{forbes&bridges10}: 92~GCs compiled from multiple previous studies that measured their ages \citep{salaris&weiss98, bellazzini+02, catelan+02, de_angeli+05, carraro+07, dotter+08, carraro09, marin_franch+09:acsgcVII},
    \item \citet{dotter+10:acsgcIX, dotter+11}: 68~GCs, whose ages were all obtained through the same method,
    \item \citet{vandenberg+13}: 54~GCs, whose absolute ages were measured consistently for the sample of \citet{marin_franch+09:acsgcVII},
    \item \citet{kruijssen+19:kraken}: 96~GCs with the mean ages determined from the three other studies.
\end{itemize}
The GC age distributions of the four studies are shown in \cref{fig:gc_observations}. The age distribution from \citet{kruijssen+19:kraken} is the one that we will also use to compare to the modelled GC ages as it takes the mean of the other studies, thereby giving us the current best absolute estimate of an individual GC's age. This is the best approach for the purposes of this study, where absolute ages are our main concern. In contrast, in many cases the relative GC ages are of greater importance, such that combining different studies should then be avoided.

\subsection{Stellar particle tracer model}

The model for GC formation by \citet{renaud+17} uses the stellar particles within a simulated galaxy as tracers for GC candidates. Only stellar particles that were formed at lookback times of more than \SI{10}{\giga\year} are considered for this and are classified as accreted and in-situ formed based on their formation sites. There is no maximum age that the GC candidates could have due to the first GC formation times still being unknown. This model was run on a cosmological hydrodynamic zoom-in simulation of a MW-like galaxy, which has a virial mass of $M_{200} = \SI{1.3e12}{\Msun}$ at $z=0$ in the parent simulation (where $M_{200}$ is the mass within the radius enclosing a volume containing a density 200 times higher than the mean density of the Universe) and a stellar mass of $M_* = \SI{4.1e10}{\Msun}$ at $z=0.5$, the last point in time of the zoom-in simulation. The last major merger of the galaxy occurred at $z \approx 2$ with a mass ratio of 1:10, similar to the predicted last major merger of the MW, the GSE \citep[e.g.,][]{belokurov+18, helmi+18}.

The modelled GC candidates statistically represent the properties of the accreted and in-situ formed clusters. At the same time, as they were directly obtained from the stellar particles, their ages also trace the overall star formation history of a simulated MW-like galaxy before \SI{10}{\giga\year} ago. The ages we used in this work are those found in the bottom panel of fig.~8 of \citet{renaud+17}. Since the formation efficiency is likely different between accreted and in-situ formed GCs, we kept the two subpopulations of GCs separate in the analysis for this work. The total population of GC candidates is then a superposition of these two.
Their used cosmological parameters are $h = 0.702$ and $\Omega_M = 0.272$.
We show the relevant ranges of galaxy and GC population properties in \cref{tab:model_galaxies} for this and the other models presented in the following.

\begin{table*}
    \centering
    \caption{Galaxy properties of the GC model studies and from the our MW: the number of MW-mass galaxies analysed ($N_\mathrm{gal}$), their virial and stellar mass ranges ($M_\mathrm{vir}$ and $M_*$), the number of GCs contained in the individual galaxies ($N_\mathrm{GC}$), the mean and standard deviation of $N_\mathrm{GC}$, and the mean and standard deviation of the ages of all the GCs.
    The virial mass definitions are $M_{200}$ for \citet{renaud+17}, \citet{kruijssen+19:emosaics}, \citet{reina_campos+22:emppathfinder}, and \citet{valenzuela+24:GCs}; total mass of the system according to the applied subhalo finder for \citet{chen&gnedin24:MW_M31}; and halo mass for \citet{de_lucia+24:GCs}.
    The GC-limited sample of \citet{de_lucia+24:GCs} is further selected according to $N_\mathrm{GC}$.
    Note that the differences in GC number and average GC age are the result of the different galaxy samples.
    For the values measured from observations, the virial mass estimate is from \citet{bobylev&baykova23}, which has also been proposed to be larger by other more recent works \citep[e.g.,][]{kravtsov&winney24}, the stellar mass from \citet{licquia&newman15}, $N_\mathrm{GC}$ from \citet{garro+24}, and we computed $\langle t_\mathrm{age} \rangle$ using the mean GC ages from \citet{kruijssen+19:kraken}.}
    \label{tab:model_galaxies}
    \begin{tabular}{llcccccc}
        \hline
        Model & Study & $N_\mathrm{gal}$ & $M_\mathrm{vir} / (\SI{e12}{\Msun})$ & $M_* / (\SI{e10}{\Msun})$ & $N_\mathrm{GC}$ & $\langle N_\mathrm{GC} \rangle$ & $\langle t_\mathrm{age} \rangle / \si{\giga\year}$ \\
		\hline
		Stellar particle tracer & \citet{renaud+17} & 1 & 1.3 & 4.1 (at $z=0.5$) & -- & -- & \num{11.4 \pm 0.7} \\
            E-MOSAICS & \citet{kruijssen+19:emosaics} & 6 & 0.7--2.2 & 0.8--4 & 60--330 & \num{130 \pm 100} & \num{10.3 \pm 2.1} \\
            EMP-Pathfinder & \citet{reina_campos+22:emppathfinder} & 21 & 0.6--2.3 & 0.4--6.5 & 140--1250 & \num{400 \pm 240} & \num{8.1 \pm 3.5} \\
            Rapid mass growth & {\citet{chen&gnedin24:MW_M31}} & 3 & 1.1--1.4 & 4--8 & 160--180 & \num{170 \pm 10} & \num{11.2 \pm 1.4} \\
            Two-phase & \citet{de_lucia+24:GCs} & \num{12884} & 0.8--3 & 2--8 & 0--\num{11100} & \num{290 \pm 590} & \num{9.5 \pm 1.7} \\
            Two-phase (GC-limited) & \citet{de_lucia+24:GCs} & \num{1140} & 0.8--3 & 2--8 & 170--250 & \num{205 \pm 25} & \num{10.8 \pm 1.1} \\
            Dual formation pathway & \citet{valenzuela+24:GCs} & 21 & 1--2 & 1.6--4.4 & 220--700 & \num{400 \pm 120} & \num{11.1 \pm 2.3} \\
		\hline
            MW Observations & & -- & \num{1.1 \pm 0.4} & \num{6.1 \pm 1.1} & ${\gtrsim}200$ & -- & \num{11.9 \pm 1.3} \\
		\hline
	\end{tabular}
\end{table*}

\subsection{E-MOSAICS model}

The E-MOSAICS model (MOdelling Star cluster population Assembly In Cosmological Simulation within EAGLE; \citealp{pfeffer+18:emosaics}) couples MOSAICS \citep{kruijssen+11:mosaics, kruijssen+12} onto the EAGLE galaxy formation model \citep{crain+15, schaye+15:eagle} as a subgrid cluster model. When gas particles form stellar particles in the simulation, a certain fraction of the formed stellar particle mass is considered to be contained in stellar clusters in a subgrid fashion depending on the local gas density, velocity dispersion, and temperature. This designated cluster mass is distributed over a cluster population according to an initial cluster mass function (ICMF) set as an exponentially cut-off power law \citep{schechter76}. Mass loss and disruption of GCs are accounted for according to the tidal forces measured throughout time. For the study of GCs, they only take the clusters with masses above $M_\mathrm{GC} > \SI{e5}{\Msun}$ and low metallicities of $-2.5 < \FeH < -0.5$ into account to get around the issue of \enquote{under-destruction} of GCs due to the multiphase ISM not being modelled. This would mostly affect younger, metal-rich GCs that spend more time orbiting the gas discs of their host galaxies.

\Citet{pfeffer+18:emosaics} and \citet{kruijssen+19:emosaics} ran the E-MOSAICS model for the volume-limited sample of 25 zoom-in cosmological simulations of present-day MW-mass galaxies contained in the \SI{25}{\mega\parsec} volume from the EAGLE project. The galaxies have virial masses of $M_{200} = \SI{7e11}{\Msun}$--\SI{3e12}{\Msun} and stellar masses of $M_* = \SI{8e9}{\Msun}$--\SI{4e10}{\Msun} at $z=0$. For this work, we extracted the ages of the GC systems within the six galaxies found in fig.~3 of \citet{kruijssen+19:emosaics}. Those galaxies cover the full range of virial and stellar masses of the parent sample of 25~galaxies and were specifically chosen by \citet{kruijssen+19:emosaics} to show the variation of GC system properties found within the whole sample of simulations.
Their used cosmological parameters are $h = 0.6777$ and $\Omega_M = 0.307$.

\subsection{EMP-Pathfinder model}
\label{sec:emp_pathfinder}

The EMP-Pathfinder model \citep{reina_campos+22:emppathfinder} addresses the issue of clusters disrupting too slowly in E-MOSAICS by including the physics of the multiphase nature of the ISM in the simulations. The shock-driven disruption of clusters by the cold gas in the ISM is necessary to reproduce the mass distribution of old GCs in the MW. The model follows that of E-MOSAICS with some changes concerning GCs, including the ICMF model, the disruption treatment, and GC size prescriptions. Mass loss and disruption of GCs are accounted for through tidal forces. Dynamical friction is applied to the clusters in post-processing.

EMP-Pathfinder was run for 21 of the MW-mass galaxies used in the E-MOSAICS project with the same zoom-in initial conditions. All six galaxies for which we have the E-MOSAICS GC data are among these galaxies. The virial masses range from $M_{200} = \SI{6e11}{\Msun}$--\SI{2.3e12}{\Msun} and stellar masses from $M_* = \SI{4e9}{\Msun}$--\SI{6.5e10}{\Msun}. While \citet{reina_campos+22:emppathfinder} focus their analysis on the clusters older than \SI{10}{\giga\year} for a large part of their work, this approach is not appropriate for our aim of studying the intrinsic predictions of GC models with respect to the GC ages. For this reason, we first set a GC mass cut of $M_\mathrm{GC} \geq \SI{e4}{\Msun}$ to be comparable to the mass range of GCs observed in the MW \citep{baumgardt&hilker18}, and then applied the same metallicity cut as \citet{kruijssen+19:emosaics} for E-MOSAICS of $-2.5 < \FeH < -0.5$. Additionally, we also consider the GC age distributions without the metallicity cut, though it should be noted that the large amount of young clusters is artificial and reflects their numerical choices for the star formation prescription.
Their used cosmological parameters are $h = 0.6777$ and $\Omega_M = 0.307$, the same as used in E-MOSAICS.

\subsection{Rapid mass growth model}
\label{sec:rapid_growth}

The GC formation model employed by \citet{chen&gnedin24:MW_M31} is based on that introduced by \citet{choksi+18}: when the relative virial mass accretion rate of a galaxy surpasses a certain threshold value (a free parameter of their model, found to be $\dot{M}_\mathrm{halo} / M_\mathrm{halo} > \SI{0.5}{\per\giga\year}$), GC formation is triggered. To obtain the available cold gas mass, analytic stellar mass-halo mass and gas mass-stellar mass relations are applied to the virial masses of the galaxies as well as prescriptions for the evolution of their scatter.
The available cold gas mass is then converted to the total formed GC mass by an efficiency factor. The number of GCs is obtained by sampling from a \citet{schechter76} ICMF. Young stellar particles or DM particles in local density peaks are then tagged with the formed GCs to trace their evolution thereafter. Mass loss and disruption of GCs are accounted for according to the tidal forces measured throughout time.

\Citet{chen&gnedin24:MW_M31} presented three MW-like galaxies with this GC model. Those galaxies were selected by their total mass, circular velocity, and accretion history, where the last major merger must have occurred between 10--\SI{12}{\giga\year} ago to resemble a GSE merger and secular evolution thereafter. Two of the MW-like galaxies are from TNG50, and one is from a DM-only zoom-in simulation of a galaxy group resembling the Local Group \citep[introduced by][]{brown&gnedin22, chen&gnedin23}. The galaxies cover mass ranges of $M_\mathrm{vir} = \SI{1.1e12}{\Msun}$--\SI{1.4e12}{\Msun} and $M_* = \SI{4e10}{\Msun}$--\SI{8e10}{\Msun} at $z=0$. For the GCs, we only take into account those with masses of $M_\mathrm{GC} \geq \SI{e4}{\Msun}$ for comparability with the observed MW GCs, just as we also did for the EMP-Pathfinder GCs.
The cosmological parameters are $h = 0.6774$ and $\Omega_M = 0.3089$ for TNG50 and $h = 0.68$ and $\Omega_M = 0.31$ for the Local Group simulation.

\subsection{Two-phase model}

\Citet{de_lucia+24:GCs} use a semi-analytic two-phase model for GC formation and evolution based on the model of \citet{kruijssen15}, where stellar clusters are formed from cold gas in the galaxy disc depending on the cold gas surface density and the angular momentum. The galaxy properties are obtained from the semi-analytic model GAEA (GAlaxy Evolution and Assembly; \citealp{hirschmann+16:gaea, xie+17:gaea}), which is run on DM-only simulations and populates the DM haloes with baryonic properties based on the merger histories. The stellar clusters are evolved through time taking into account mass loss and disruption until a galaxy merger is able to eject the clusters into the halo. Note that as a semi-analytic model, the actual GC spatial properties are not traced. Only the clusters that the model ejects into the halo as a consequence of a merger are considered to be GCs.

The model was run alongside GAEA for the Millenium Simulation \citep{springel+05:millennium}, a DM-only cosmological simulation with a box length of \SI{685}{\mega\parsec}. They define their MW-like galaxy selection as being disc-dominated (bulge-to-total mass ratio $B/T < 0.2$) and that are central galaxies with virial masses of $M_\mathrm{vir} = \SI{8e11}{\Msun}$--\SI{3e12}{\Msun} at $z=0$. To make the galaxy selection more comparable with those of the other studies selected in this work, we further add a stellar mass criterion of $M_* = \SI{2e10}{\Msun}$--\SI{8e10}{\Msun}, leading to a total of \num{12884} galaxies.
For the GCs, we only take into account those with masses above $M_\mathrm{GC} \geq \SI{e4}{\Msun}$, just as for some of the previously presented models. No metallicity cut is applied.
As the number of GCs in those galaxies varies greatly (see \cref{tab:model_galaxies}), we also consider a second more limited sample of galaxies, taking only those into account with 170--250~GCs (this is the case for 1140 galaxies, whereas \num{8146} galaxies have fewer GCs and \num{3598} have more).
This number range is based on the estimated number of GCs found in the MW of 170 with measured kinematics \citep{vasiliev&baumgardt21} and more than 200 including newly identified GCs in the bulge region \citep{garro+24}.
Their used cosmological parameters are $h = 0.73$ and $\Omega_M = 0.25$.

\subsection{Dual formation pathway model}

The model by \citet{valenzuela+21} employs two GC formation pathways, where GCs are either formed in low-mass galaxies when the virial mass surpasses a certain threshold value of around \SI{6e8}{\Msun} or when a galaxy experiences a large accretion rate. The second formation pathway follows the rapid growth model introduced by \citet{choksi+18} and on which the model by \citet{chen&gnedin24:MW_M31} is based, see \cref{sec:rapid_growth}. The galaxy properties are obtained from the empirical galaxy formation model \code{EMERGE} \citep{moster+18:emerge}. No GC mass loss or disruption is accounted for, therefore the model traces only the formation of surviving GCs. Since we only analyse the ages of GCs having survived until $z=0$, this is not an issue for this work.

The model was run alongside \code{EMERGE} for a DM-only cosmological simulation of side length \SI{30}{\mega\parsec}. The MW-mass galaxies were selected as the 21~central galaxies with virial masses of $M_\mathrm{vir} = \SI{1e12}{\Msun}$--\SI{2e12}{\Msun}, of which two were identified as having a similar GC age distribution to the MW, with one having an accretion history that resembles an early GSE-like merger followed by no significant accretions \citep{valenzuela23, valenzuela+24:GCs}. The MW-mass galaxies have stellar masses ranging $M_* = \SI{1.6e10}{\Msun}$--\SI{4.4e10}{\Msun}.
Their used cosmological parameters are $h = 0.6781$ and $\Omega_M = 0.308$.

\subsection{Other models}
\label{sec:other_models}

There are some GC models that do not create GCs according to a physical formation prescription, but place them into galaxies according to unrelated criteria that assume an already existing GC population in a particular circumstance. Even though these models are tuned to GC properties at $z=0$, they therefore do not predict GC formation times, but rather inform about other GC properties. We mention two such models in the following, as well as one that was only very recently introduced.

The model by \citet{boylan_kolchin17} tests what the average host mass of GCs at $z=6$ needs to be to reproduce the numbers of GCs in galaxies at $z=0$, assuming that the surviving GCs already exist at $z=6$. For this reason, their model simply places all GCs in the galaxies at $z=6$, and can thus not be used to predict GC ages.

The GC model by \citet{doppel+23} follows that of \citet{ramos_almendares+20} and places GCs in satellite galaxies at the time of accretion onto a more massive host galaxy by tagging DM particles, thus following the general spatial and orbital properties of the infalling satellite galaxy. This means that the model does not use any prescription to physically form GCs and thereby the GC \enquote{seeding times} only trace the GC infall time onto the host galaxy rather than the actual GC formation. The model therefore does not predict any GC ages.
They apply the model to the 39 most massive groups and clusters in the cosmological simulation TNG50. Two of those systems are also among the objects classified as MW analogues by \citet{pillepich+24}. Those are galaxies with stellar discs, stellar masses of around $M_* = \SI{3e10}{\Msun}$--\SI{1.5e11}{\Msun}, and have no other massive galaxies above $10^{10.5}\,\Msun$ within a \SI{500}{\kilo\parsec} environment.
For completeness, we show the GC seeding times as tracers of their infall times onto the main galaxy in \cref{app:doppel_gcs}.

Finally, a very recent addition to GC models is presented by \citet{chen+24:GCs}, where GCs are formed in subclouds that appear from fragmentation of turbulent gas when dark matter haloes accrete matter rapidly at high redshifts. Two populations of GCs are formed through two channels, the first by compression by gravity or supernova feedback, the second by cooling in metal-poor subclouds, which results in an older metal-poor and a younger metal-rich GC population. In this work, we discuss the results from the studied models with those presented by \citet{chen+24:GCs} when possible.

\section{GC age distributions for MW-like galaxies}
\label{sec:age_distributions}

In the following we compare the GC age predictions of the different models with each other and with observations. For the models which have been applied to a large sample of galaxies, we here focus on those galaxies with the most similar GC age distributions as the observed GCs in the MW.

\subsection{Modelled GC age distributions}

\begin{figure*}
    \centering
    \includegraphics[width=0.9\textwidth]{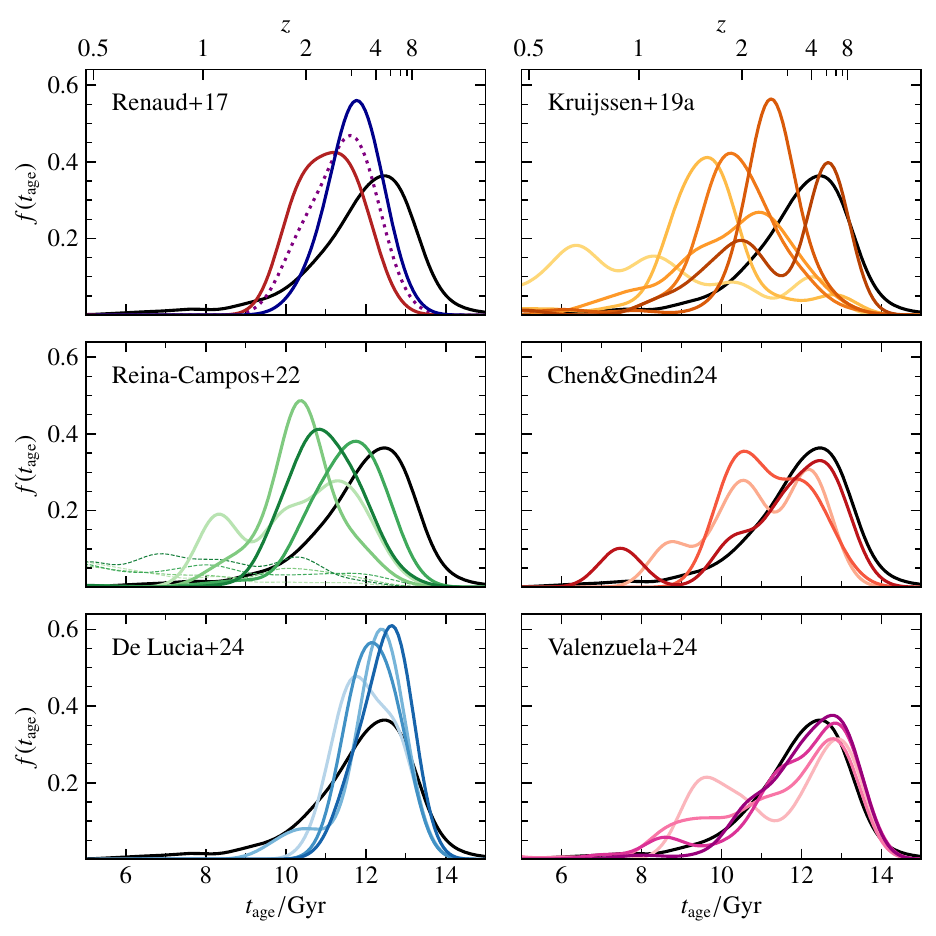}
    \caption{Age distributions of modelled GCs in MW analogues compared to the observed GC ages in the MW.
    The full age distributions reaching to $t_\mathrm{age} = \SI{0}{\giga\year}$ are shown in \cref{fig:gc_model_panels_young}.
    The observed GC ages are the mean values taken from \citet{kruijssen+19:kraken} and are shown in each panel as the black line. The modelled GC ages are from \citet{renaud+17}, \citet{kruijssen+19:emosaics}, \citet{reina_campos+22:emppathfinder}, \citet{chen&gnedin24:MW_M31}, \citet{de_lucia+24:GCs}, and \citet{valenzuela+24:GCs}. The dotted blackberry line for \citet{renaud+17} is the combined age distribution of the in-situ (red) and accreted (blue) GC age distributions, assuming an equal number of underlying GCs for each group.
    For the other models the colours are selected to be darker for galaxies with older mean GC ages.
    The dashed lines for \citet{reina_campos+22:emppathfinder} correspond to the same galaxies as the solid distributions of equal colour, with the difference that no metallicity cut was applied to the GCs, which results in a large population of young star clusters.
    This is best seen in \cref{fig:gc_model_panels_young}.
    The age distributions are determined from the sum of normal distributions at the individual age measurements with standard deviations equal to the individual age uncertainties. For the modelled GCs, we assume a uniform age uncertainty of \SI{0.5}{\giga\year}.}
    \label{fig:gc_model_panels}
\end{figure*}

For each of the GC models introduced in the previous section, we show the modelled GC age distributions for some of the MW-like galaxies the models were applied to in \cref{fig:gc_model_panels}, always including the mean observed MW GC ages from \citet{kruijssen+19:kraken} for reference and better comparability (black line in each panel).
To visualise the differences of the typically old GC ages the best, the age axis is restricted to values between $t_\mathrm{age} = \SI{5}{\giga\year}$--\SI{15}{\giga\year}. As not all the studies only have GC ages in that range, we additionally show the full age distributions in \cref{app:young_ages}.

In the top left panel, the in-situ (red) and accreted (blue) GC age distributions of \citet{renaud+17} are shown, as well as the total GC age distribution (dotted), assuming equal numbers of in-situ and accreted GCs.
In the top right panel, the GC age distributions are shown of all six E-MOSAICS galaxies from fig.~3 of \citet{kruijssen+19:emosaics}.
In the middle left panel, the four galaxies with the on average oldest GCs are shown from EMP-Pathfinder \citep{reina_campos+22:emppathfinder} with solid lines, where the metallicity cut as described in \cref{sec:emp_pathfinder} is applied. The age distributions of all star clusters without an additional metallicity cut are shown with dotted lines. This leads to the inclusion of a large population of metal-rich clusters that are young.
In the middle right panel, the three MW-like galaxies from \citet{chen&gnedin24:MW_M31} are shown. The distribution in the lightest colour is from the Local Universe simulation, the two darker ones from TNG50.
In the bottom left panel, four galaxies from the GC-limited sample by \citet{de_lucia+24:GCs} are shown, meaning they have 170--250~GCs. The four galaxies were selected from the galaxies with the oldest GC populations.
Finally, in the bottom right panel, four galaxies from \citet{valenzuela+24:GCs} are shown, two of which were selected by the authors as having similar GC age distributions as the MW (the two lighter colours), and two additional ones with old GC populations (the two darker colours).

Here we focus first on these galaxies selected to be similar to the MW in terms of their GC age distributions, but some of the models have GC ages available for more MW-mass galaxies that may have significantly different assembly histories and morphologies than our MW. We present and discuss their varieties of GC age distributions as predictions for other MW-mass galaxies in \cref{sec:age_distribution_variance} for the models where this is possible.
Before discussing these differences in more detail, we would like to reiterate what the aim is of this side-by-side comparison, and what it is not.
All of the studies define a sample of MW-mass or MW-like simulated galaxies, for which they compare the respective GC populations to that of our MW. For this reason, the benchmark for these systems is the MW itself and our aim in this work is to lay out the predicted GC ages side by side of the current GC models for the MW-like galaxies.
However, as each of the studies applied their GC model to one or multiple simulated galaxies or generated merger trees, and only the galaxies from E-MOSAICS and EMP-Pathfinder \citep{kruijssen+19:emosaics, reina_campos+22:emppathfinder} overlap, the shown age distributions are results of potentially entirely different assembly histories and are subject to a whole range of baryonic physics and galaxy formation models. Furthermore, each study selects MW-mass/MW-like or MW analogue galaxies according to their own criteria, and different definitions concerning the resulting GC populations are used, such as the mass or metallicity cuts, which can in particular affect their age distribution.
As a consequence, we do not claim to be able to fully interpret the differences found between the models as such differences will arise from a combination of factors that all play a role.

All the considered GC models are capable of producing GCs for certain MW-like galaxies that have overall old GCs ($t_\mathrm{age} \gtrsim \SI{8}{\giga\year}$), just as is the case for the MW. To zeroth order, this means that these GC models are consistent with each other and with observations. Of course, having six models each with their own approach to GC formation, differences concerning their predicted GC formation times are not surprising as the details surrounding GC formation are still poorly understood. These differences can also be directly seen in \cref{fig:gc_model_panels}. Additionally, the observed GC ages have large systematic uncertainties on the order of \SI{1}{\giga\year}--\SI{2}{\giga\year}, such that the observed age distribution has to be taken with caution. However, there is in general consensus about the MW GCs being particularly old, most likely as a result of the early formation history \citep[e.g.,][]{helmi+18}.

\subsection{Differences between the models}

While the model by \citet{renaud+17} simply assumes any stellar particle older than \SI{10}{\giga\year} to be a potential host of a GC, this means that their age distributions in fact traces the early star formation history of the MW-like galaxy until $t_\mathrm{age} = \SI{10}{\giga\year}$. Clearly, the accreted (blue) stars are on average slightly older than the in-situ formed stars (red), by ${\sim}\SI{1}{\giga\year}$. Still, the observed MW GCs are measured to be even older than the accreted stars, again by around $\SI{1}{\giga\year}$ for the peak formation times. This shows that even the peak early star formation before $t_\mathrm{age} = \SI{10}{\giga\year}$ in their simulated MW-like galaxy occurs later than the peak GC formation in our MW. Other studies have also determined the GCs being older than the field stars \citep[e.g.,][]{harris&harris02, peng+08, reina_campos+19}. Assuming this prediction from their simulation is correct and considering the findings that the cosmic star formation peaked around $z \sim 2$ \citep{madau&dickinson14,driver+18}, this means that the formation of surviving GCs does not follow the early star formation history ($z > 2$) of a galaxy, but occurs earlier.
The large systematic uncertainties of the observed ages could again bring the simulated and observed distributions into agreement, however.

For the E-MOSAICS and EMP-Pathfinder models, the GC ages tend to be younger than found in the MW, with the exception of the oldest E-MOSAICS GC system, which also appears to experience a second GC formation burst at around $t_\mathrm{age} \sim \SI{10}{\giga\year}$. While this at first may look similar to the result of an early merger like the GSE, those younger GCs were actually formed in another galaxy that merged much later in time \citep{kruijssen+19:emosaics}. The older measured MW GCs could be an indication that none of the 25 selected E-MOSAICS galaxies has an accretion history that is as early as that of the MW. However, for both E-MOSAICS and EMP-Pathfinder it is necessary to apply additional cuts to the metallicity to exclude especially metal-poor and metal-rich GCs, where the latter GCs are also mostly young. This cut is applied to differentiate the GCs from young/open clusters in those models. The effect of not applying such a cut can be seen for the EMP-Pathfinder GCs from the dashed lines. Being the only two subgrid models implemented within hydrodynamical cosmological simulations, this implies that there are still open questions concerning the exact GC formation conditions and their disruption. In particular for GC disruption, major advances were already made between E-MOSAICS and EMP-Pathfinder by including the cold ISM, but this appears to not have strongly affected the first formation times of GCs.

Despite being the study with the by far largest sample of galaxies, even when only considering the galaxies with 170--250~GCs, the oldest GC age distributions from \citet{de_lucia+24:GCs} barely reach the high observed GC ages in the MW. This is also mentioned by the authors, who state that the GC systems are systematically younger than the observed MW GCs by ${\sim}\SI{1}{\giga\year}$, though again the systematic uncertainties of the observations could be a part of the reason. Interestingly, the spread of GC ages is much smaller than for most of the other models except for \citet{renaud+17}. This is likely the result of individual merger events producing many GCs at once by ejecting clusters from the disc into the halo, or because of a coarse time binning, which would lead to the appearance of many GCs forming at once. In \cref{sec:age_distribution_variance} we present further GC age distributions of different mean ages from \citet{de_lucia+24:GCs} and discuss why it is more likely that individual merger events are the cause for most of their GC formation.

\Citet{chen&gnedin24:MW_M31} selected three MW-like galaxies from two simulation suites using multiple criteria based on the properties and formation history of the MW to identify best-matching MW analogues. One of their galaxies has an especially old GC population with an age distribution similar to that of the MW (dark red), while both of the other galaxies have a more significant GC formation peak at $t_\mathrm{age} = \SI{9}{\giga\year}$--\SI{11}{\giga\year}, which roughly corresponds to the time range in which at least one major merger was enforced to exist to mimic the infall of GSE. Such a massive merger at early times will have been gas-rich and thus is expected to form a large population of GCs in the rapid mass growth model of \citet{chen&gnedin24:MW_M31}, as predicted by \citet{valenzuela+24:GCs} using the same model assumptions.

Similarly, the selected galaxies from \citet{valenzuela+24:GCs} show old GC ages consistent with the measured ones from the MW. These also feature secondary formation peaks between $t_\mathrm{age} \sim \SI{8}{\giga\year}$ and \SI{11}{\giga\year}, indicating major merger events around that time. As their model includes a second formation pathway beyond the one employed by \citet{chen&gnedin24:MW_M31}, which populates dwarf galaxies in low-mass haloes with GCs at early times, it is not surprising that the GCs are slightly older than those of \citet{chen&gnedin24:MW_M31} and that the secondary peaks at GSE infall times are less pronounced compared to the oldest GC population.

\subsection{Empirical and subgrid GC models}

Interestingly, the empirical models (i.e., models where the GC formation prescription is based on empirical or semi-analytic prescriptions, regardless of whether particle tagging is applied) tend to form the oldest GC populations. While \citet{kruijssen+19:emosaics} find one galaxy in E-MOSAICS with an especially early burst of GC formation, that galaxy only has around 60~GCs in total with masses above $M_\mathrm{GC} \geq \SI{e5}{\Msun}$ and also has a later GC formation peak with a longer tail towards younger ages ending at $t_\mathrm{age} \sim \SI{8}{\giga\year}$. While this does not affect the relative distribution of ages, it still means that none of the E-MOSAICS galaxies actually forms a large absolute number of GCs earlier than $t_\mathrm{age} \sim \SI{12}{\giga\year}$, which is also the case for EMP-Pathfinder.
The reason for the empirical models overall forming the earlier GCs is likely related to them being independent of the implemented baryonic physics in hydrodynamical cosmological simulations as they assume observationally-based scaling relations for the baryonic galaxy properties, such as star formation rate, cold gas mass, and stellar mass. As cosmological simulations have had issues with reproducing observed early star bursts, relying on these prescriptions for GC formation may be leading to GCs forming at the wrong times, even if the local physical formation conditions were perfectly understood.

On the other hand, the empirical models of \citet{chen&gnedin24:MW_M31} and \citet{valenzuela+24:GCs} do not follow the self-consistent formation of galaxies within their DM haloes, but assume average scaling relations to apply. As a result, the GC populations of individual galaxies have to be considered as realisations based on correct galaxy merger histories, but with average baryonic properties. However, these scaling relations and the approach of forming GCs in gas-rich systems with rapid mass growth lead to very old GCs, as is expected from observations.
Nevertheless, the disadvantage of using average scaling relations is that galaxies will overall have little gas at later times, therefore making late GC formation much rarer in galaxies, regardless of the individual assembly histories. This means that galaxies falling outside the norm with large gas amounts will not be directly covered by these models, in particular in the model by \citet{valenzuela+21, valenzuela+24:GCs}, who directly employ the scaling relation for the gas depletion time without taking into account its scatter, whereas the stellar mass-halo mass relation is intrinsically dealt with in the underlying empirical galaxy formation model \code{EMERGE}.
\Citet{chen&gnedin24:MW_M31} include prescriptions for the scatter of the scaling relations, among them that of \citet{behroozi+13:sfh} for the stellar mass-halo mass relation. In particular for the stellar mass-halo mass relation, a large scatter may be necessary to obtain the amount of bright galaxies observed with JWST at high redshifts \citep[e.g.,][]{mason+23, mirocha&furlanetto23, shen+23}. Increasing this scatter would lead to a larger population of more massive galaxies, which in turn could give birth to more massive stellar clusters at formation time and modify the resulting age distribution of surviving GCs. This issue is in part circumvented in the model by \citet{valenzuela+24:GCs} through the free parameter of the formation pathway in low-mass galaxies.

The model by \citet{de_lucia+24:GCs} takes the middle ground between the subgrid and empirical models: implemented within a semi-analytic galaxy formation model, baryonic properties are inferred for the DM haloes through the semi-analytic model based on the merger histories. The model thus does not spatially resolve the internal gas behaviour, just like the empirical models. At the same time, it applies a similar GC formation prescription as employed in E-MOSAICS, being based on the approach of \citet{kruijssen15}.

As small time differences in the early Universe correspond to much larger differences in redshift than at later times, the variations of predicted GC formation ages in fact imply significantly different epochs of GC formation in cosmological terms. With the ongoing observations of heavily star-forming galaxies at high redshifts with JWST \citep[e.g.,][]{claeyssens+23, adamo+24, mowla+24}, narrowing down our understanding of GC formation will be vital to compare the predictions of improved models with the observed systems as well as to find to what extent massive star clusters at high redshifts are the progenitors of today's GCs. This will be further discussed in \cref{sec:redshift_distributions}.

\section{GC age distributions for other MW-mass galaxies}
\label{sec:age_distribution_variance}

While some of the considered studies only have GC data available for a small number of MW-like galaxies that we could fully show in \cref{fig:gc_model_panels}, we have data available for a larger sample of MW-mass galaxies from EMP-Pathfinder \citep{reina_campos+22:emppathfinder}, \citet{de_lucia+24:GCs}, and \citet{valenzuela+24:GCs}. These studies can therefore be used as a broader prediction to other MW-mass galaxies. In the case of \citet{de_lucia+24:GCs}, this applies to MW-mass disc galaxies, whereas the galaxies from EMP-Pathfinder and \citet{valenzuela+24:GCs} are MW-mass galaxies of all types.

\begin{figure}
    \centering
    \includegraphics[width=0.9\columnwidth]{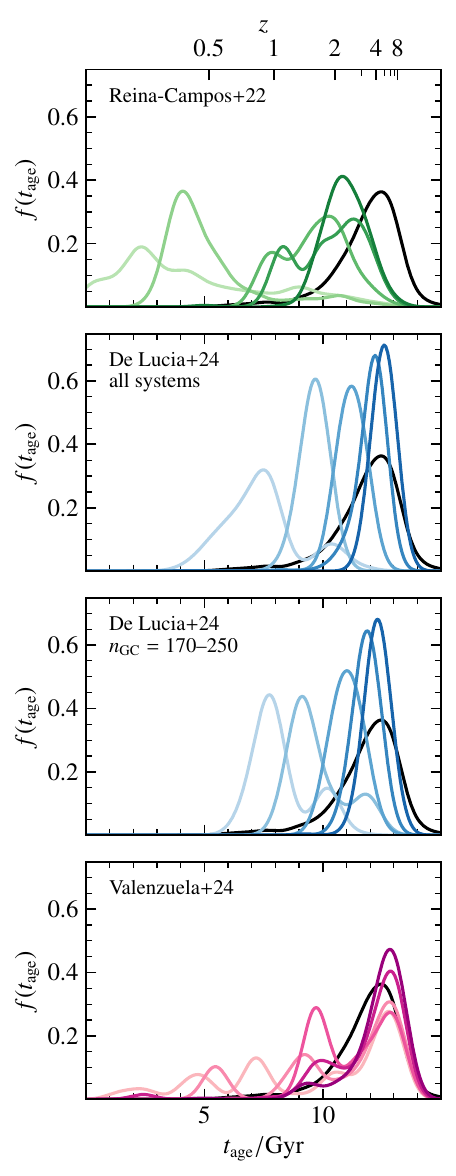}
    \caption{Age distributions of modelled GCs in MW-mass galaxies compared to the observed GC ages in the MW. The observed GC ages are the mean values taken from \citet{kruijssen+19:kraken} and are shown in each panel as the black line. The modelled GC ages are from \citet{reina_campos+22:emppathfinder}, \citet{de_lucia+24:GCs} when considering all MW-like galaxies (second row) and when considering only galaxies with 170--250~GCs (third row), and \citet{valenzuela+24:GCs}. For each study, five example galaxies are shown: computing the mean GC age for each galaxy, here the median galaxy and the $1\sigma$ and $2\sigma$ outlier galaxies hosting younger or older GCs are shown. The colours are chosen to be darker for older GC populations.}
    \label{fig:gc_model_distribution_panels}
\end{figure}

In \cref{fig:gc_model_distribution_panels} we show the whole range of predicted GC age distributions for MW-mass galaxies from EMP-Pathfinder, \citet{de_lucia+24:GCs}, and \citet{valenzuela+24:GCs}. For \citet{de_lucia+24:GCs}, we show the distributions considering all their selected galaxies (second row), and the distributions only considering the selected galaxies with 170--250~GCs contained in them (third row; see \cref{tab:model_galaxies} for the properties of the galaxies and GC systems). Sorting the galaxies by their mean GC ages, we plotted the median and the $1\sigma$ and $2\sigma$ younger and older GC populations from each study, with darker colours indicating the older systems.

The first striking difference is the range that the age distributions of the three models cover. The youngest GC populations in EMP-Pathfinder include a significant fraction of GCs having ages smaller than $t_\mathrm{age} \sim \SI{5}{\giga\year}$ and even \SI{3}{\giga\year}, and the median to oldest GC populations have peak formation times of $t_\mathrm{age} \sim \SI{10}{\giga\year}$ to \SI{12}{\giga\year}. This indicates that most of their MW-mass galaxies have relatively old GC populations, but the younger half of the GC populations extends far into the late times of the Universe.
Similarly, the six galaxies from E-MOSAICS seen in \cref{fig:gc_model_panels} span a variety of different ages, with most of the GC populations having relatively old ages ($t_\mathrm{age} \gtrsim \SI{10}{\giga\year}$). However, due to the GC data only being available for six galaxies, we cannot draw further conclusions from the spread of age distributions.

The model from \citet{de_lucia+24:GCs} has peak formation times between $t_\mathrm{age} \sim \SI{7}{\giga\year}$ and \SI{13}{\giga\year}. The five selected GC age distributions shown for the GC-limited galaxy sample (third row) have formation peaks relatively evenly distributed with time, indicating that the resulting mean GC ages of the galaxies are roughly normally distributed, with a slightly longer tail towards younger ages. It can also be seen that the spread of GC formation is larger the later the GCs are formed, also resulting in a lower relative age peak (as each age distribution is normalised). This would mean that GCs are formed over shorter periods of time in the early Universe. Besides two of the galaxies with constrained GC number having minor secondary GC formation peaks (the two with the younger GC populations), all of the shown galaxies have age distributions dominated by a single GC formation peak.
For the full galaxy sample, the GC ages span the same range as for the GC-limited sample. In both galaxy samples the peaked GC ages imply that the formation of surviving GCs is actually tied to single events that create GCs in a very short period of time, as was already suggested in \cref{sec:age_distributions}.
This is likely the result of merger events leading to the ejection of clusters from the disc into the halo, where they are considered as GCs in the model. Clusters are subject to mass loss and disruption in the disc before they are ejected, such that mergers will mostly only eject clusters having formed shortly before the event while older clusters would have already been disrupted for the most part.

In contrast to the first two studies, the model from \citet{valenzuela+24:GCs} always produces a dominant old population of GCs with ages of $t_\mathrm{age} > \SI{11}{\giga\year}$. The reason for this old population in their model is the formation pathway of forming GCs in small haloes as soon as they surpass a certain threshold virial mass. As this mass is rather low and in the regime of dwarf galaxies, this leads to significant GC formation at these early times. As these small haloes are the building blocks of the assembling more massive galaxies, they provide the proto-MW-mass galaxies with old GCs. Most of the younger GCs are then formed through the second formation pathway of rapid accretion onto the central galaxy, leading to the younger bumps of the GC age distribution seen for the five galaxies in the bottom panel of \cref{fig:gc_model_distribution_panels}.

As can seen by the vastly different possible GC age distributions predicted by the various models, there is still a large disagreement concerning how young the GC population of an individual MW-mass galaxy can actually be. While the studies by \citet{kruijssen+19:emosaics}, \citet{reina_campos+22:emppathfinder}, and \citet{de_lucia+24:GCs} predict the existence of some MW-mass galaxies having no or only very few old GCs and to varying degrees how young an entire GC population can be, the model from \citet{valenzuela+24:GCs} always contains a significant fraction of old GCs. On the even more extreme side, \citet{boylan_kolchin17} assumed that all GCs had been formed by $z=6$ for their model. Traditionally, it has been assumed that the bulk of GCs are old with ages of $t_\mathrm{age} \gtrsim \SI{10}{\giga\year}$ \citep{cote+98, strader+05} based on the findings in our own MW and on more GC age estimates in nearby galaxies (e.g., \citealp{usher+19, usher+24}; Cabrera-Ziri et al.\ in prep.). Some galaxies also have been found to feature young massive clusters (YMCs), which have properties that are consistent with those expected for GCs right after formation time \citep[e.g.,][]{beasley+02}. Of the galaxies that have been studied for these purposes, \ngc{3377} stands out as a peculiar galaxy having a large young population of GCs \citep{usher+19}. However, even this early-type galaxy is measured to have many old GCs. The question thus stands whether MW-mass galaxies with purely younger GCs or YMCs exist, such as has been observed for certain lower-mass galaxies like the SMC \citep[e.g.,][]{parisi+14}. More estimates of GC ages in other MW-mass galaxies will be needed to discern between the models. For this, even rough estimates of the ages will be sufficient, which may even be possible through photometric age determinations. This will especially be relevant for the recently launched and upcoming large area low surface brightness surveys like Euclid, Rubin, and Roman.

As an important point to keep in mind, we want to highlight that the definitions of GCs vary across models. Some differences are directly given by the mass-cuts intrinsically used by the models or applied by us to the GC data, such as the lower-mass cut: the GCs we show for E-MOSAICS have $M_\mathrm{GC} \geq \SI{e5}{\Msun}$, for EMP-Pathfinder and \citet{chen&gnedin24:MW_M31} $M_\mathrm{GC} \geq \SI{e4}{\Msun}$, and the GCs from the model of \citet{valenzuela+24:GCs} have initial masses of $M_\mathrm{GC,init} \geq \SI{e5}{\Msun}$. Metallicity cuts to the E-MOSAICS and EMP-Pathfinder remove metal-rich GCs with mostly young ages that are assumed to be artificial due to under-disruption and an overproduction of clusters at $z<1$, respectively. Finally, there is only a fuzzy line between what would be identified as a GC, a YMC, or an open cluster from observations as well as from models. It is therefore also difficult to tell in what way the GC output of a model would have to be compared with YMCs or even open clusters in addition to GCs.

Clearly, the currently available GC models have very different predictions concerning the variety of GC age distributions in MW-mass galaxies. These deviations are significant enough that the differences in galaxy masses, morphologies, or in the definitions of GCs will not be sufficient to bring all the predictions in agreement. To summarise the range that the models cover: some of them predict that every single MW-mass galaxy should have a dominant old GC population with $t_\mathrm{age} > \SI{10}{\giga\year}$ \citep[e.g.,][]{valenzuela+24:GCs}, whereas others predict that roughly a third of MW-mass disc galaxies have most of their GCs being younger than \SI{5}{\giga\year} \citep[e.g.,][]{reina_campos+22:emppathfinder}.
Having investigated the overall ages of the GCs in the models, we turn to the high-redshift formation times of surviving GCs from the same set of models in the next section.

\section{Predictions for high-redshift clusters}
\label{sec:redshift_distributions}

\begin{figure*}
    \centering
    \includegraphics[width=0.9\textwidth]{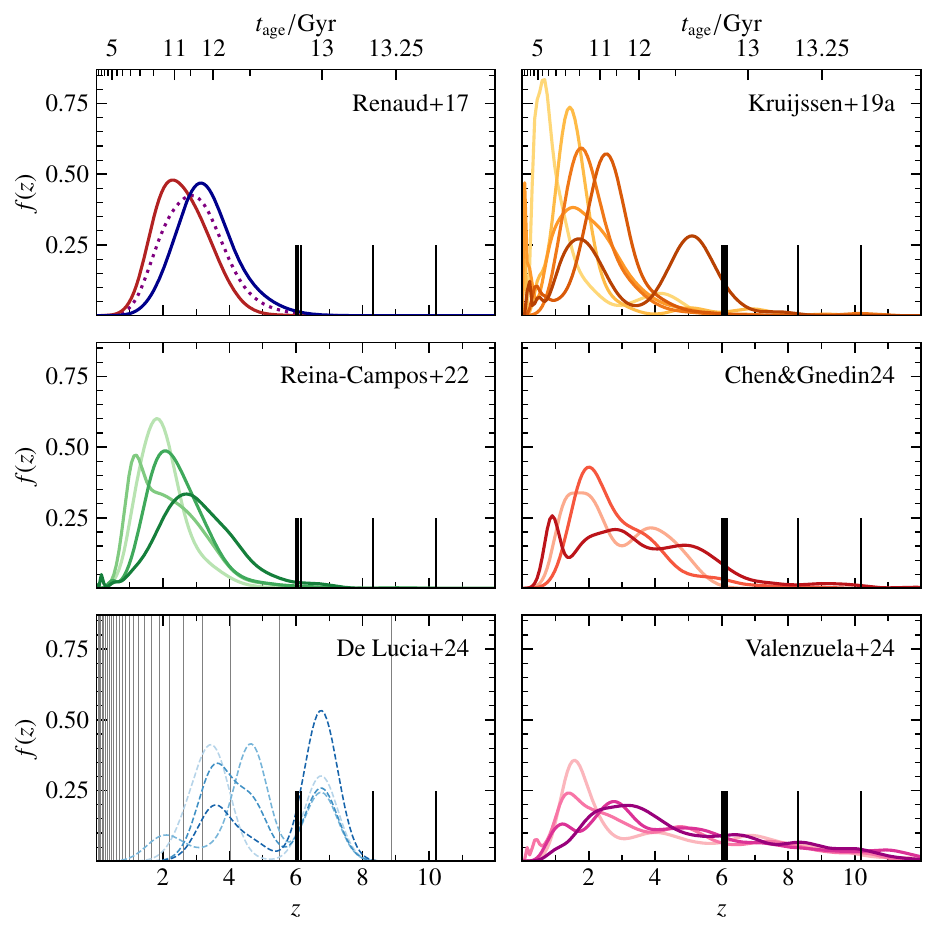}
    \caption{Formation redshift distributions of surviving modelled GCs in MW analogues.
    The formation redshifts are from \citet{renaud+17}, \citet{kruijssen+19:emosaics}, \citet{reina_campos+22:emppathfinder}, \citet{chen&gnedin24:MW_M31}, \citet{de_lucia+24:GCs}, and \citet{valenzuela+24:GCs}.
    The short vertical black lines denote the redshifts of star clusters from the strongly lensed systems observed by \citet{vanzella+23}, \citet{fujimoto+24}, \citet{messa+24:lyalpha}, \citet{mowla+24}, and \citet{adamo+24}, in increasing redshift order.
    The dotted blackberry curve for \citet{renaud+17} is the combined age distribution of the in-situ (red) and accreted (blue) GC age distributions, assuming an equal number of underlying GCs for each group.
    As the GC ages for the galaxies from \citet{de_lucia+24:GCs} are only available in equally-spaced age bins that are relatively wide at high redshifts, the curves are plotted with dashes and the bin edges are indicated as vertical grey lines, with the last bin edge reaching a redshift of $z \approx 8.9$.
    For all the models, the formation redshift distributions are determined from the sum of normal distributions at the individual formation redshift with standard deviations equal to the individual redshift itself until $z=0.5$, and beyond that equal to 0.5. This prevents the smoothed curves from surpassing $z=0$ by too much in the plots.}
    \label{fig:gc_model_panels_redshift}
\end{figure*}

For the recent and ongoing observations of clustered star formation in high-redshift galaxies with HST \citep[e.g.,][]{messa+24} and JWST \citep[e.g.,][]{claeyssens+23, vanzella+23, adamo+24, mowla+24}, we test in the following what predictions can be made from the GC models with respect to such star clusters in high-redshift potential progenitors of MW-mass galaxies.
In \cref{fig:gc_model_panels_redshift}, we show the GC formation time distributions of the same set of galaxies from each of the six GC models as shown in \cref{fig:gc_model_panels}, just with the redshift plotted uniformly along the x-axis instead of age. For this, we converted the ages to the corresponding formation redshifts according to the original cosmologies applied by the respective authors (see \cref{sec:data}). For the GCs by \citet{de_lucia+24:GCs}, the ages were only available with equal-width age bins, such that the distribution becomes more discrete in redshift-space at high redshifts. For this reason, the lines are dashed and we included the age bin edges in grey for reference.

The redshifts of some of the high-redshift star clusters found with JWST are marked by short vertical black lines, which include the Sunrise arc ($z=6.0$; \citealp{vanzella+23}), the Cosmic Grapes ($z=6.072$; \citealp{fujimoto+24}), MACS J0416 ($z=6.143$; \citealp{messa+24:lyalpha}), the Firefly Sparkle ($z=8.304$; \citealp{mowla+24}), and the Cosmic Gems arc ($z=10.2$; \citealp{adamo+24, bradley+24}).
Their host galaxies are expected to grow to be MW-mass galaxies.
For an overview of the current high-redshift observations of star clusters, see \citet{pfeffer+24:highz}.

Taking the formation time distribution from \citet{renaud+17} as tracer for the overall star formation history of a MW-like galaxy, it can be seen that the surviving GCs produced by the other models generally start forming at earlier times than the rest of the stars.
All models have similar predictions with respect to the peak formation redshifts for the surviving GC populations: these lie around $z=1$--3 and vary from galaxy to galaxy. This corresponds to the observed formation time of young massive clusters at $z=2.4$ in the Sunburst galaxy \citep{vanzella+22:sunburst} and it is slightly after current observations at $z=4$-5 of young massive clusters \citep[e.g.,][]{vanzella+22:abell2744, messa+24}.
Only the oldest galaxies from the sample of \citet{de_lucia+24:GCs} all have no young GC population and have peak formation redshifts somewhere between $z=3$ and $z=8$. Most of the GC systems in their model do not have surviving GCs being formed before $z=6$, however. Also \citet{chen+24:GCs} show a peak age at $z=5$ (see their fig.~9). Additionally, one of the E-MOSAICS galaxies has two similarly dominant formation peaks at $z \approx 2$ and at $z \approx 5$.
While the peak formation time is therefore quite consistent among the models, the spread to higher redshifts varies significantly: the overall star formation as traced by the model of \citet{renaud+17} begins at $z=5$ and $z=6$ for the in-situ and accreted components, respectively. This is also the case for EMP-Pathfinder, with one of the systems even reaching surviving GC formation times as early as $z=7$. E-MOSAICS shows a similar behaviour for most of their galaxies, though one in particular has an especially early surviving GC formation burst starting at $z=7$--8.
For the high-redshift clusters formed in E-MOSAICS, \citet{pfeffer+24:highz} also compares the brightest cluster properties with recent JWST and HST observations.

The model by \citet{chen&gnedin24:MW_M31} produces more surviving GCs at these larger redshifts, but only start their formation slightly earlier, at $z=6$--10 for the very first GCs of some of their galaxies. The formation tail towards high redshifts is even longer for \citet{valenzuela+24:GCs}, where for the shown MW-like galaxies the surviving GC formation rate per redshift slowly increases almost linearly, starting at around $z=12$.
For these models that use empirical scaling relations, increasing the scatter of the halo mass-stellar mass relation could lead to more massive galaxies and star clusters in lower-mass halos, resulting in earlier formed surviving GCs. This approach may be necessary to find better agreement with JWST observations of bright galaxies at high redshifts of $z \gtrsim 9$ \citep[e.g.,][]{mason+23, shen+23}.
The initial formation time of surviving GCs in the model by \citet{chen+24:GCs} is even slightly earlier at $z=14$ (see their fig.~9).
Finally, the only information at higher redshifts we can extract from the binned data of \citet{de_lucia+24:GCs} is that there is a non-significant number of surviving GCs formed in the range of $z=5.5$--9, with none being formed before $z=9$. However, as these are among the oldest GC systems of almost 700 galaxies, this prediction is not valid for the majority of their MW-mass galaxies. Linking this to the high-redshift observations marked with the vertical black lines, clearly the model that predicts that the star clusters at $z>8$ could very well survive until today is the one by \citet{valenzuela+24:GCs}. With a much lower probability of this happening, its is principally also possible according to E-MOSAICS and the model by \citet{chen&gnedin24:MW_M31}.

The models with younger surviving GC populations at $z=0$ are therefore either inconsistent with these observational findings, or they predict that almost none of the high-redshift star clusters would turn into a GC at present day, as this would require more of the oldest GCs having formation times at higher redshifts. The models with the older GC populations (\citealp{chen&gnedin24:MW_M31, valenzuela+24:GCs}; and for certain galaxies \citealp{kruijssen+19:emosaics, de_lucia+24:GCs}), in contrast, find a larger number of GCs formed at high redshifts ($z \gtrsim 6$), thereby suggesting that a larger fraction of the high-redshift star clusters could in fact be the progenitors of today's GCs around MW-mass galaxies.

High-resolution simulations of the early Universe have been able to trace the formation of potential proto-GCs at high redshifts. For instance, \citet{phipps+20} list the properties of several stellar clusters having been formed in the cosmological simulation FiBY (First Billion Years) before $z=6$, though the identified clusters all have masses below $\SI{3e5}{\Msun}$, which is at and below the lower end of the inferred cluster masses from observations at those redshifts \citep[e.g.,][]{messa+24, mowla+24}. It is therefore even less likely than for the observed clusters that the lower-mass systems identified in the simulation would survive to become GCs at $z=0$.
\Citet{mayer+24} find stellar clusters forming at $z>7$ through disc fragmentation and discover a very similar mass surface density distribution of their objects compared to those observed at similar redshifts \citep[e.g.,][]{mowla+24, adamo+24}.
The stellar clusters formed through high-resolution proto-galaxy mergers in the FIRE simulation \citep{kim+18} and in dwarf galaxy mergers \citep{lahen+19} are found to have masses up to $\SI{e6}{\Msun}$ \citep{lahen+19}, consistent with some of the observed high-redshift clusters.

One central question is how many of high-redshift galaxies actually contain massive star clusters, as current observations are still rather limited in number. Combining such a value with the GC models run for a diverse set of galaxies will be able to predict the surviving star cluster fraction as a function of redshift.
Observations of systems at intermediate redshifts already indicate that their GCs must have formed at very early times of $z = 7$--11, as found by \citet{mowla+22} for the Sparkler at $z \approx 1.4$, though also with significant systematic age uncertainties, which strongly affect the corresponding formation redshifts. This indicates that some of the early formed massive bound star clusters survive until at least intermediate redshifts.

More observations at high redshifts and further discriminating between the current GC models will bring further insight into the connection of GCs and star clusters and clumps at high and low redshifts, including what fraction of the observed high-redshift clumps would be expected to survive to present day, and if this depends on the mass or type of the galaxies in question.

\section{Conclusion}
\label{sec:conclusion}

In this work, we present the first side-by-side comparison of multiple current GC formation models, for which we collected GC age data from the GC models applied to MW-mass and MW-like galaxies to compare their age distributions with each other and with observations of the GCs in the MW as well as observations of star clusters at high redshifts.
We give a detailed overview of the models and their methods for forming and evolving GCs side-by-side alongside their individual definitions for galaxies being MW-like and for the GCs themselves.
Among the models are subgrid models applied to zoom-in hydrodynamical cosmological simulations \citep{kruijssen+19:emosaics, reina_campos+22:emppathfinder}, empirical models applied to merger trees of cosmological simulations \citep{chen&gnedin24:MW_M31, valenzuela+24:GCs}, and GC models implemented within semi-analytic models of galaxy formation \citep{de_lucia+24:GCs}, as well as a simple tagging prescription of old stellar particles in a hydrodynamical simulation \citep{renaud+17}.
The objective of this direct comparison of the model results is to present in a clear way the range of predictions that current GC models make, which lays the foundation for future development and improvement of such models.

For a selection of MW-like galaxies, we find that all of the considered models are in general capable of forming GCs at ages older than \SI{10}{\giga\year} as is observed for the MW. While the model by \citet{renaud+17} can be used as a tracer of the regular star formation history at early times, the current best estimates for the MW GC ages indicate that GCs are formed earlier than the overall stellar population. While it should be noted that the MW GC absolute ages have large uncertainties, the E-MOSAICS \citep{kruijssen+19:emosaics} and EMP-Pathfinder \citep{reina_campos+22:emppathfinder} galaxies show slightly younger age distributions offset by 1--\SI{3}{\giga\year} depending on the simulated galaxy. Only one old E-MOSAICS galaxy is directly consistent with the best age estimates for the MW GCs. For a carefully selected sample of MW-like galaxies according to their merger histories, \citet{chen&gnedin24:MW_M31} find very similar age distributions. \Citet{valenzuela+24:GCs} find the overall oldest GC age distributions, which are also similar to those observed. Finally, the oldest systems of the large sample by \citet{de_lucia+24:GCs} have peak surviving GC formation times consistent with the MW.

An important point to keep in mind for this is always that the different MW-like galaxies can have very different accretion histories, thus preventing an apples-to-apples comparison of the ages.
For this reason, we also compared the variety of age distributions expected for MW-mass galaxies based on the larger galaxy samples from \citet{reina_campos+22:emppathfinder}, \citet{de_lucia+24:GCs}, and \citet{valenzuela+24:GCs}. While the model \citet{valenzuela+24:GCs} predicts that every galaxy has a significant old GC population, the GC systems from \citet{de_lucia+24:GCs} show a roughly normal distribution of peak GC formation times, with half of their galaxies having most GCs being younger than 10--\SI{11}{\giga\year}. The model by \citet{reina_campos+22:emppathfinder} even predicts that roughly a third of MW-mass disc galaxies have most GCs being younger than \SI{5}{\giga\year} and lack an old GC population.
Discerning between these models will therefore be possible by estimating rough ages of GCs in nearby galaxies. This means that it is critical that further observations of GCs around other galaxies in the MW mass range be performed and ages be estimated for them, such as what has been done for \m{31} \citep[e.g.,][]{usher+24}. Recent and upcoming large area low surface brightness surveys like Euclid, Rubin, and Roman will provide the means necessary to perform such rough measurements to increase the statistics of GC ages.

Finally, we compared the formation redshifts of surviving GCs from the models to the observed formation times of star clusters in high-redshift galaxies from HST and JWST. The models consistently predict a peak formation redshift of $z=1$--3, depending on the individual galaxy. However, the tail towards higher redshifts varies significantly. Some models begin producing the first surviving GCs at $z=5$--7 \citep{renaud+17, kruijssen+19:emosaics, reina_campos+22:emppathfinder}, while others produce surviving GCs even earlier at $z=6$--10 \citep{chen&gnedin24:MW_M31} and $z=12$ \citep{valenzuela+24:GCs}.
Compared to recent observations of star clusters in strongly-lensed galaxies at $z=6$--10 \citep[e.g.,][]{vanzella+23, mowla+24, adamo+24}, the models with older GC populations predict that at least some of these systems could survive until present day, whereas other models are either inconsistent with these observations or predict that such systems are always disrupted over cosmic time.
Increasing the sample of high-redshift star clusters in observations will be of great value for combining these with GC models to obtain predictions for the chances of survival and thus statistically quantifying how many of such star clusters are in fact proto-GCs.

In the future, this type of model comparison will be extended to further GC properties to test to what extent and for which use cases each of the models can be appropriately applied. To overcome the limitations of comparing GCs obtained within galaxies with different properties and accretion histories, this should ideally be carried out by running each of the models for the same galaxy initial conditions.

\section*{Acknowledgements}

We thank Yingtian (Bill) Chen, Gabriella De Lucia, Oleg Gnedin, Marta Reina-Campos, Florent Renaud for helpful discussions and for providing us with their modelled GC data or making them publicly available.
We also thank Joel Pfeffer and the anonymous referee for useful comments that have improved the paper.
LMV acknowledges support by the German Academic Scholarship Foundation (Studienstiftung des deutschen Volkes) and the Marianne-Plehn-Program of the Elite Network of Bavaria.
DAF thanks ARC for its support via DP220101863 and DP200102574.
This research was supported by the Excellence Cluster ORIGINS, funded by the Deutsche Forschungsgemeinschaft under Germany's Excellence Strategy -- EXC-2094-390783311.
The following software was used for this work: Julia \citep{bezanson+17:julia}, CSV.jl \citep{quinn+:csv.jl}, DataFrames.jl \citep{kaminski+:dataframes.jl}, and \code{matplotlib} \citep{hunter07:matplotlib}.

\section*{Data Availability}

The data underlying this article will be shared on reasonable request to the corresponding author.



\bibliographystyle{style/mnras}
\bibliography{bib} 




\appendix

\section{Globular cluster formation time distributions of other models}
\label{app:doppel_gcs}

\begin{figure}
    \centering
    \includegraphics[width=\columnwidth]{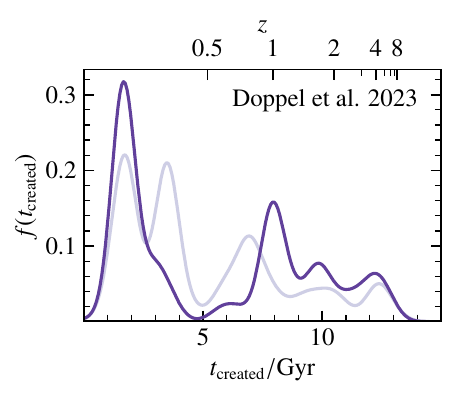}
    \caption{Seeding time ($t_\mathrm{created}$) distributions of the modelled GCs in two simulated MW analogue galaxies in TNG50 from \citet{doppel+23}.}
    \label{fig:gc_model_doppel}
\end{figure}

As the GC seeding times from the model by \citet{doppel+23} are not physical formation times due to the GCs being seeded at the time of infall of satellite galaxies (see \cref{sec:other_models}), we did not include these times in the main analysis of this work. For completion, we show the distribution of GC seeding times for the two TNG50 MW analogue galaxies for which the GC model was run in \cref{fig:gc_model_doppel}. This is effectively the model's prediction for the time of GC accretion. The seeding times are overall much later than the observed and modelled GC ages as seen in \cref{fig:gc_model_panels}. While some GCs appear to be accreted earlier than \SI{10}{\giga\year} ago, the majority of GCs have been accreted within the last ${\sim}\SI{3}{\giga\year}$. This indicates that these MW analogues have had a different accretion history than our MW, as the majority of observed MW GCs have either been formed in-situ in the early formation of the Galaxy or were accreted with GSE around \SI{8}{\giga\year}--\SI{11}{\giga\year} ago.

\begin{figure*}
    \centering
    \includegraphics[width=0.9\textwidth]{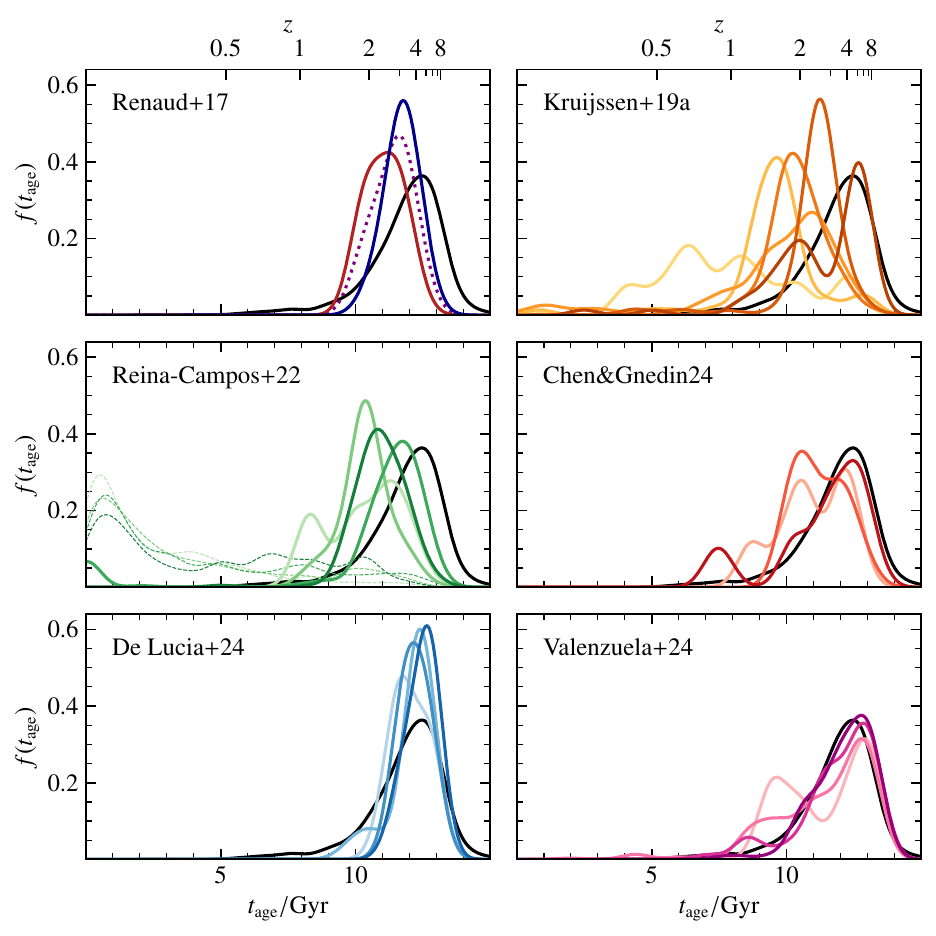}
    \caption{Age distributions of modelled GCs in MW analogues compared to the observed GC ages in the MW. This plot shows the same GC age distributions as in \cref{fig:gc_model_panels}, but with ages extending to values of $t_\mathrm{age} = \SI{0}{\giga\year}$. The dashed lines for EMP-Pathfinder correspond to the same galaxies as the solid distributions of equal colour, with the difference that no metallicity cut was applied to the GCs, which results in a large population of young stellar clusters.}
    \label{fig:gc_model_panels_young}
\end{figure*}

The question still arises in what way this model also includes in-situ formed GCs, as these would be expected either to be formed in-situ in the main progenitor branch of the merger tree and thus not ever be accreted, or to be formed in another galaxy and \enquote{accreted} at very early times, thereby effectively belonging to the in-situ GC population. The overall lack of early accreted GCs indicates that either these galaxies have had much later accretion histories compared to the MW, leading to a very small early in-situ GC component, or that in-situ GCs are for the most part missing in the model.

\section{Ages including young GCs}
\label{app:young_ages}

As some of the models like E-MOSAICS \citep{kruijssen+19:emosaics} and EMP-Pathfinder \citep{reina_campos+22:emppathfinder} include significant amounts of GCs younger than \SI{5}{\giga\year}, we show the full GC age distributions from \cref{fig:gc_model_panels} with ages reaching until the present day in \cref{fig:gc_model_panels_young}. The selected host galaxies are the same as those in \cref{fig:gc_model_panels}.


\bsp	
\label{lastpage}
\end{document}